\documentclass[conference]{IEEEtran}
\IEEEoverridecommandlockouts
\usepackage{cite}
\usepackage{amsmath,amssymb,amsfonts}
\usepackage{mathtools}
\usepackage{algorithm,algpseudocode} 
\usepackage{multicol}
\usepackage{graphicx}
\usepackage{textcomp}
\usepackage{xcolor}
\usepackage{booktabs,multicol,caption}
\usepackage{url}

\DeclareMathOperator*{\argmax}{argmax}

\def\BibTeX{{\rm B\kern-.05em{\sc i\kern-.025em b}\kern-.08em
    T\kern-.1667em\lower.7ex\hbox{E}\kern-.125emX}}
\begin{document}

\title{Maximizing Clearance Rate by Penalizing Redundant Task Assignment in Mobile Crowdsensing Auctions}

\author{Maggie E. Gendy\textsuperscript{1}, Ahmad Al-Kabbany\textsuperscript{1,2,3}, and Ehab F. Badran\textsuperscript{1} \\
	\textsuperscript{1} Department of Electronics and Communications Engineering,\\
	 Arab Academy for Science, Technology, and Maritime Transport, Alexandria, Egypt \\ 
	 \textsuperscript{2}Intelligent Systems Lab, Arab Academy for Science, Technology, and Maritime Transport, Alexandria, Egypt \\ 
	\textsuperscript{3} Department of Research and Development, VRapeutic, Alexandria, Egypt
}

\maketitle

\begin{abstract}
This research is concerned with the effectiveness of auctions-based task assignment and management in centralized, participatory Mobile Crowdsensing (MCS) systems. During auctions, sensing tasks are matched with participants based on bids and incentives that are provided by the participants and the platform respectively. Recent literature addressed several challenges in auctions including untruthful bidding and malicious participants. Our recent work started addressing another challenge, namely, the maximization of clearance rate (CR) in sensing campaigns, i.e., the percentage of the accomplished sensing tasks. In this research, we propose a new objective function for matching tasks with participants, in order to achieve CR-maximized, reputation-aware auctions. Particularly, we penalize redundant task assignment, where a task is assigned to multiple participants, which can consume the budget unnecessarily. We observe that the less the bidders on a certain task, the higher the priority it should be assigned, to get accomplished. Hence, we introduce a new factor, the \emph{task redundancy factor} in managing auctions. Through extensive simulations under varying conditions of sensing campaigns, and given a fixed budget, we show that penalizing redundancy (giving higher priority to \emph{unpopular tasks}) yields significant CR increases of approximately $50\%$, compared to the highest clearance rates in the recent literature.

\end{abstract}


\section{Introduction}
\label{sec:intro}

Mobile crowdsensing (MCS) is an emerging paradigm in which sensor-rich, smart, \emph{online} mobile devices are used to accomplish various sensing tasks. These tasks are requested by service demanders from participants who are willing to share information and sensed data, through a coordinating platform. MCS systems have been deployed in diverse fields such as environmental applications \cite{b5} (e.g., measuring air quality and noise levels), infrastructure research \cite{b6} (e.g. measuring traffic congestion and road conditions), and social applications \cite{b7} (e.g. share restaurant information and crowd counting). Due to their support to a wide variety of applications, centralized participatory sensing is widely applied in MCS systems. This framework consists of a central platform and a number of smartphones, and requires active involvement from users in sensing and decision-making. Recently proposed MCS applications and platforms include MOSDEN \cite{b1}, Campaignr\cite{b3}, and Medusa\cite{b4}. These platforms are used mainly for task publishing and data collection. 

In MCS, one of the critical limitations is the restrained computational and power resources of the sensing smartphones. Hence, most participants need a kind of compensation in order to participate. High-paced research has focused on developing effective incentive mechanisms (including monetary and non-monetary approaches \cite{b44}) in order to ensure the users' willingness to share their sensing data. In relation to the scope of this research, we limit the discussion to monetary mechanisms. This implies that the coordinating platform should have a budget. The authors of \cite{b43} defined two incentive mechanisms: 1) The platform-centric model, where the static payment for each winner is determined by the platform. 2) The user-centric model, that is a reverse auction approach, where the platform has no control over payments to each winner.

\begin{figure}[t]
	\begin{center}
		\setlength{\abovecaptionskip}{3pt plus 0pt minus 0pt}
		\setlength{\belowcaptionskip}{-15pt plus 0pt minus 0pt}
		\includegraphics[width=3.4in]{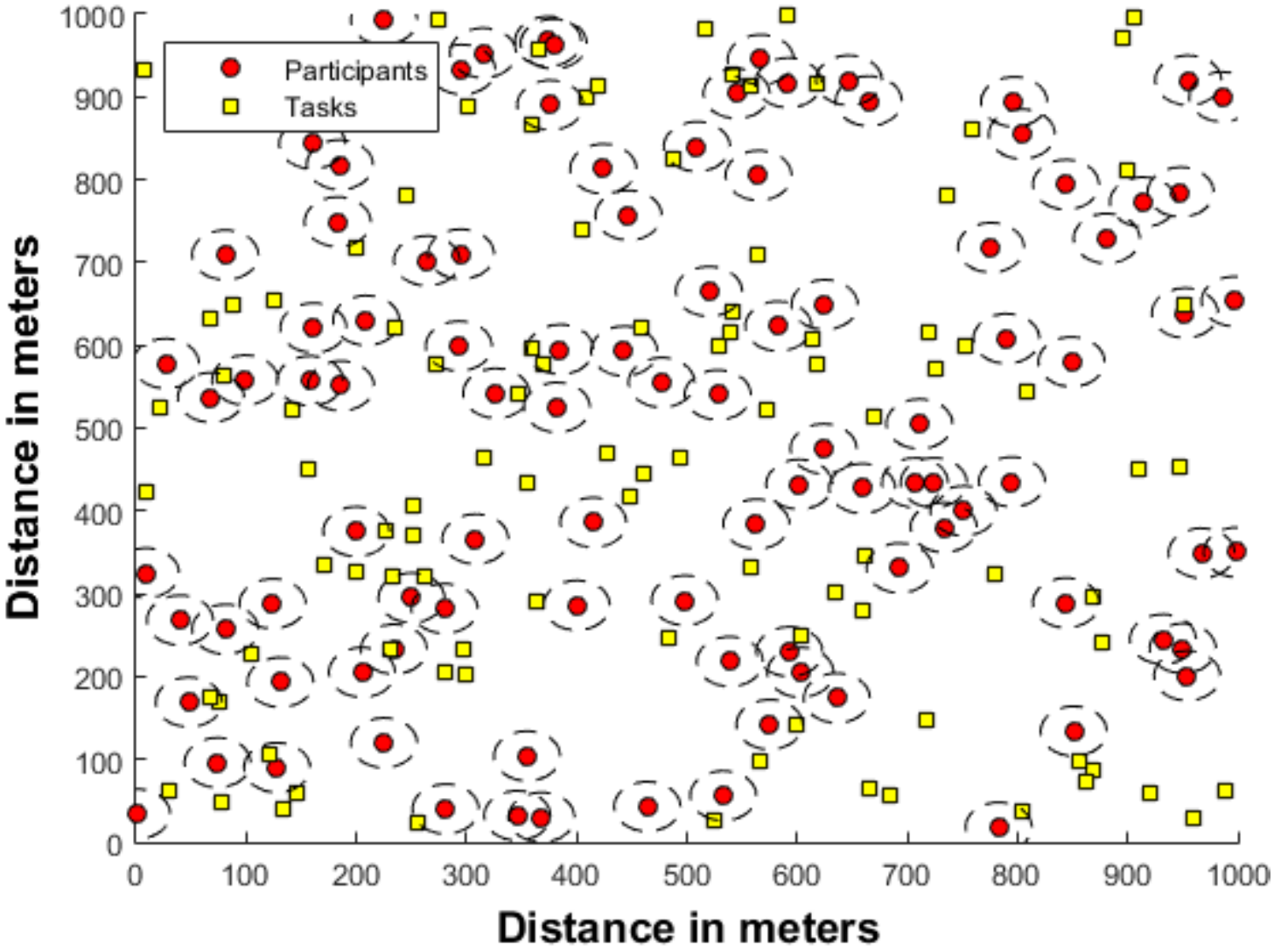}
		\caption{An illustration inspired by \cite{b43} depicting participants (red dots) interested in tasks (yellow squares) within their areas of interest (dashed circles).}
		\label{fig:ParticipantTask}
	\end{center}
\end{figure}

Among the main challenges of MCS systems is the participant selection (or similarly, the task allocation). Task allocation techniques can be classified using diverse approaches, one of which is the single-task \cite{b8, b9} vs. multi-task approach \cite{b14}. For the former family of methods, the platform selects the best set of winners to perform a task, each of whom is required to complete one task while preserving some kind of constraint (e.g., a budget, satisfying probabilistic coverage constraints, etc.). In \cite{b13}, for example, a framework was proposed to select an optimal set of winners while satisfying budget constraints. The goal of \cite{b15}, however, was to select participants such that they maximize the spatial coverage of crowdsensing. Fig.~\ref{fig:ParticipantTask} depicts a geographical area in which participants and tasks are uniformly distributed, and each participant is surrounded by an area of interest, out of which, that participant does not bid on any tasks.

Another classification approach for the task allocation process in MCS is discussed in \cite{b16}, where the authors presented a three-model perspective of task allocation. One of these models is \emph{the participant model} which is comprised of three kinds of participant traits (attributes, requirements, and supplements). 
Most importantly, the attributes indicate the inherent characteristics of participants and whether they are calculated by the platform or provided by the users. In this context, the authors of \cite{b17} proposed a reputation-aware (RA) algorithm for task allocation. Moreover, requirements from the participants side such as privacy and energy efficiency were addressed in \cite{b18} and \cite{b19} respectively. However, none of the aforementioned studies considered the campaign tasks characteristics.

While significant research attention has been given to improving the task allocation and incentive mechanisms as two important stages in MCS applications, much less attention has addressed the maximization of the clearance rate (CR), i.e., the number of accomplished tasks in auction-based campaigns. \emph{The motivation for seeking high CR is that it corroborates the satisfaction of service demanders, since they accomplish more tasks while preserving the existing auction constraints}. Hence, CR is directly proportional to high platform's utility and efficiency. To the best of our knowledge, \textbf{\emph{our own recent work}} in \cite{b50,b51} was the first to propose CR-maximized auctions by using new bidding procedures that helped increasing the CR significantly. 

\begin{figure}[t]
	\begin{center}
		\setlength{\belowcaptionskip}{-15pt plus 0pt minus 0pt}
		\includegraphics[width=3.4in]{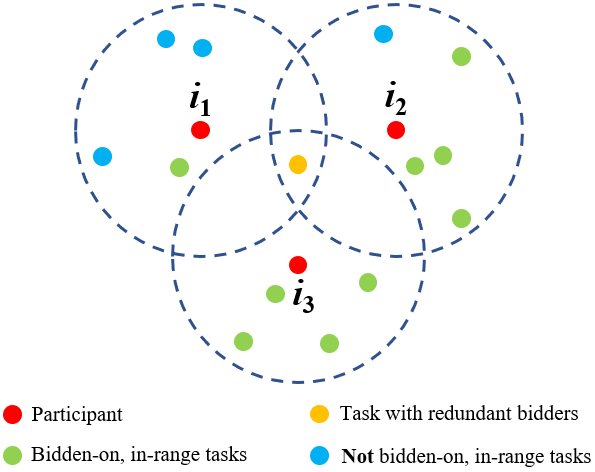}
		\caption{An illustration of a task (yellow circle) with redundant bidders $i_1$, $i_2$, and $i_3$. Green circles symbolize bidden-on tasks within the range of interest of the three participants. Blue circles symbolize tasks that are not bidden-on, yet lie within the range of interest of the three participants.}
		\label{fig:RedundancyIllustration}
	\end{center}
\end{figure}

In this research, we pinpoint one principal reason for budget drainage, and consequently not attaining satisfactory clearance rates, namely, redundant task assignment. Figure~\ref{fig:RedundancyIllustration} depicts a case of redundant task assignment where multiple participants simultaneously bid on a particular task. The figure also depicts cases for bidden-on tasks that are within the range of interest of the corresponding participants, and other cases for tasks that are not bidden-on even though they lie within the range of interest of the participants. The range of interest for each participant is shown as a dashed circle with a red circle (the participant) at its center. In the rest of this paper, we use the terms \emph{clearance rate}, \emph{task completion ratio}, and \emph{task coverage ratio} interchangeably.

The contributions of this research can be summarized as follows:
\begin{enumerate}
\item We propose a new bidding-based, multi-task allocation procedure for maximizing the platform utility by maximizing the number of covered tasks in a campaign. This is done by a new objective function that assigns a higher priority to tasks that will be completed by a fewer number of bidders--penalizing redundant task assignment.

\item To the best of our knowledge, this research is the first to use redundant task assignment in formulating a new bidding procedure that maximizes the CR of auction-based reputation-aware MCS systems.

\item Through extensive simulations, we demonstrate a remarkable enhancement in task completion ratios compared to other methods in the recent literature.

\item Similar to \cite{b50,b51}, our proposed bidding procedure links the number of campaign tasks to the platform budget, which is closer to real-life scenarios than neglecting budget constraints \cite{b17} or assuming a constant-yet-arbitrary budget \cite{b25}.
\end{enumerate}

\section{Auctions Based on Redundancy-penalizing Bidding}
\label{sec:proposed}
In this section, we propose and discuss a new bidding procedure, namely, the \emph {Redundancy-penalizing Bidding (RPB)} algorithm. Towards the goal of maximizing the CR of auctions, \emph{RPB} is based on the observation that tasks with fewer bidders should be assigned a higher priority. This increases the odds of accomplishing those tasks eventually. 
\subsection{System Model}
\label{subsec:SysMod}
Similar to previous work in the literature, the proposed bidding procedure is a greedy algorithm that is an approximation of the NP-hard problems of task allocation and auction winner selection \cite{b17, b43, b50,b51}. The steps of the \emph{RPB} algorithm are given by algorithm listing~\ref{algo:GetRedWin}, that is preceded by algorithm listing~\ref{algo:GetPrimWin}, and followed by algorithm listings~\ref{algo:GetWinPay} and \ref{algo:SecWinSecPay} \cite{b50,b51}. The whole pipeline is comprised of the following four stages, on which we elaborate below: 1) Primary Winners Selection, 2) Redundancy Winners Selection, 3) Winners Payment Determination, 4) Secondary Winners Selection and Payment. Symbols and notations that are used in the aforementioned algorithm listings are given in Table~\ref{tab1}.

For every sensing campaign, there is a crowd of participants, $\mathcal{I}$, and each smartphone $i \in \{ 1,\cdots,N \}$ represents a participant in the auction. The platform sends the details of the $M$ campaign tasks, where tasks are indexed by $j \in \{1,\cdots,M\}$. All of the participants should take part in the bidding process, and each bidder should bid on at least one task within their area of interest.

\subsection{Collective and Descriptive Bidding}
\label{subsec:Biddings}
The research in \cite{b50,b51} highlighted the difference between two types of bidding, namely, \emph{collective bidding} and \emph{descriptive bidding}. The former is the classical form of bidding, commonly discussed in the literature, that resembles a wholesale or bidding in bulk, where user $i$ asks for one collective payment in return for all the tasks in $T_i \subset T$. The set $T$ is comprised of all the tasks in a sensing campaign, i.e., $|T|=M$, while $T_i$ is the subset of tasks in which the participant $i$ is interested, i.e., $|T_i|=M_i$. In descriptive bidding, however, a participant sends a list of tasks and a separate bid for each of them. Throughout this document, we refer to this list as \emph{the list of per-task user bids}. The summation of the per-task user bids for the user $i$ is given by:
\begin{equation}
b_i = \sum_{j=1}^{M_i}x_{ij} \times b_{ij}, 
\label{eqn:sumbids}
\end{equation}
where $b_{ij}$ is the bid of user $i$ for task $j$ (descriptive bid), and $x_{ij}$ is a binary flag indicating if the user $i$ is interested in task $j$ or not. Unless the bidder is interested in only one task, the sum of the descriptive bids is usually more than the collective bid. 

\subsection{Primary Winners Selection}
\label{subsec:Primary}
Following previous work in the literature \cite{b17}, the proposed pipeline starts by calculating the marginal contribution (or marginal value) for each participant, as formulated by \cite{b17}, and then subtracts their collective bids from the resultant value (line 3 in algorithm listing~\ref{algo:GetPrimWin}). In this formulation, namely, the reputation-aware (\emph{RA}) formulation, the collective bid of user $i$, $b^c_i$, is weighted by the user's reputation score $r_i$, such that a high reputation score would result in lowering the bid, and consequently increases the odds of selecting that user. Afterwards, tasks are allocated to a set of winners, $S$, named primary winners, (lines 5,6 in algorithm listing~\ref{algo:GetPrimWin}), that are chosen such that the budget (same as platform utility in \cite{b17}) is preserved.

\begin{table*}[t]
	\centering
	\begin{center}
		\caption{Symbols and notations used in the algorithm listings}
		\label{tab1}
	\begin{tabular}{p{1cm}|p{7cm}||p{1cm}|p{6.5cm}}
		\toprule
		\textbf{Symbol} & \textbf{Meaning} & \textbf{Symbol} & \textbf{Meaning} \\ \hline
		$v_j$ & Value of task $j$ & $V$ & Set of values of tasks \\
		\hline
		\addlinespace[0.25em]
		 $v_i^{u}(S)$ & Reputational-Redundant value for user $i$ over set $S$ & $T_S$  & Set of tasks done by users in set $S$ \\ \hline
		\addlinespace[0.25em]
		$v^r_i(S)$  & Reputational value for user $i$ over set $S$ & $B^c$ & Set of collective bids by participants  \\ \hline
		\addlinespace[0.25em]
		$T^s_i(S^s)$ & Set of tasks allocated for secondary winner $i$ over the set $S^s$ & $p_i$ & Payment to winner $i$ \\ \hline
		\addlinespace[0.25em]
		$R$ & Set of participants' reputation values  & $P$ & Set of payments to participants
		\\ 
		\bottomrule
	\end{tabular}
    \end{center}
\end{table*}

\begin{algorithm}
	\caption{Determining Primary Winners}
	\label{algo:GetPrimWin}
		\begin{algorithmic}[1]
			\Function{Get Primary Winners}{$V$, $B^c$, $R$, $P$}
    				\State {$S \leftarrow \Phi$}, {$S_{tasks} \leftarrow \Phi$}
				\vspace{0.25em}
    				\State $h \leftarrow \argmax\limits_{i \in \mathcal{I}}{v_i^r(S)-\frac{b_i^c}{r_i}}$
				\vspace{0.25em}
				 \While {$\frac{b_h^c}{r_h} < v_h^r \hspace{2pt} \bigwedge \hspace{2pt} S \neq P$}
					\vspace{0.25em}
				 	\State $S \leftarrow S \cup {h}$ \hspace{2pt},\hspace{2pt} $S_{tasks} \leftarrow T_h$
					\vspace{0.25em}
					\State $h \leftarrow \argmax\limits_{i \in \mathcal{I} \setminus S}{v_i^r(S)-\frac{b_i^c}{r_i}}$
					\vspace{0.25em}
				\EndWhile
				\State \Return $S$
			\EndFunction
			\end{algorithmic}
\end{algorithm}

\subsection{Redundancy Winners Selection}
\label{subsec:RedWinSel}
This sub-section highlights the main contribution of this research. Various aspects may lead to the easiness of accomplishing certain tasks. e.g., their location, as illustrated in Fig.~\ref{fig:ParticipantTask}. Hence, many participants might bid on them. Although impactful on the CR, and hence on the utility of the platform, as shown in the next section, unpopular tasks have not been addressed efficiently by the previously proposed bidding methods. \textbf{\emph{In this paper, we introduce a new redundancy factor}} that is given by: 
\begin{equation}
d_i =1- max_{t \in T_i} \{ \frac {1}{|\Gamma_t|} \},
\label{eqn:tasks}
\end{equation}
where $d_i$ is the redundancy factor of user $i$ (such that $D$ is the set of values for participants' redundancy factor), and $|\Gamma_t|$ is the cardinality of the set of participants who are bidding on the task $t$. 

We need to increase the opportunity of user $i$ (of being selected) if $i$ is interested in a task $t \in T_i$ for which there are a few bidders, i.e., if $d_i$ is small. Hence, the more participants bidding on a task, the less priority it gets. Towards this goal, the platform adopts a procedure that is similar to the primary winners selection procedure. Particularly, in order to choose the set of redundancy winners $S^{R}$, the platform uses a weighted version of the reputation score, as given by algorithm listing~\ref{algo:GetRedWin}. This \emph{weighted reputation score} is named the \emph{redundancy-reputation factor} and is given as
\begin{equation}
u_i = \frac {r_i}{d_i},
\label{eqn:Ufactor}
\end{equation}
where $r_i$ is the reputation of user $i$ (such that $R$ is the set of participants' reputation values), and $u_i$ is the redundancy-reputation factor of user $i$ (such that $U$ is the set of values of participants' redundancy-reputation factor). The higher the $u_i$, the higher the opportunity of user $i$ to be selected as a winner in the auction. The main idea proposed by this research (penalizing redundant task assignment) can be seen in the objective functions in lines $4$ \& $7$ of algorithm listing~\ref{algo:GetRedWin}. The significant impact of the redundancy-reputation factor on the attained clearance rates will be shown and discussed in the Results and Discussion section. It is worth mentioning that for reputation-unaware (RU) bidding \cite{b17}, $r_i=1$ for user $i$.

\begin{algorithm}
	\caption{Identifying Redundancy Winners}
	\label{algo:GetRedWin}
		\begin{algorithmic}[1]
			\State  $Compute \hspace{0.1cm} U \hspace{0.1cm} for\hspace{0.1cm} participants$
                        \Function{Get Redundancy Winners}{$V$, $B^c$, $U$, $P$}
    				\State {$S^{R} \leftarrow \Phi$}
				\vspace{0.25em}
    				\State $h \leftarrow \argmax\limits_{i \in \mathcal{I}}{v_i^{u}(S^{R})-\frac{b_i^c}{u_i}}$
				\vspace{0.25em}
				 \While {$\frac{b_h^c}{u_h} < v_h^{u} \hspace{2pt} \bigwedge \hspace{2pt}S^{R} \neq P$}
					\vspace{0.25em}
				 	\State $S^{R} \leftarrow S^{R} \cup {h}$ \hspace{2pt},\hspace{2pt} $S_{tasks} \leftarrow T_h$
					\vspace{0.25em}
					\State $h \leftarrow \argmax\limits_{i \in \mathcal{I} \setminus S^{R}}{v_i^{u}(S^{R})-\frac{b_i^c}{u_i}}$
					\vspace{0.25em}
				\EndWhile
				\State {$S^{R}= S^{R}\setminus (S^{R} \cap S)$}
			    \State \Return $S^{R}$
			\EndFunction
			\end{algorithmic}
\end{algorithm}

The platforms proceeds with the payment calculation for both sets of winners, namely, the primary winners and the redundancy winners. This is accomplished using the procedure given in algorithm listing~\ref{algo:GetWinPay}. The algorithm is comprised of two consecutive loops; the first loop computes the payments of the primary winners, and the second loop computes the payments of the redundancy winners. Each of these loops is similar to the payment computation procedure that is given in \cite{b50,b51}. 

Penalizing redundant task assignment results in saving the platform's budget, since the mathematical expression of the budget $\mathcal{B}$ is given by:
\begin{equation}
\mathcal{B} = \mathcal{V}-\mathcal{P},
\label{eqn:bpv}
\end{equation}
where $\mathcal{V}$ is the sum of values of campaign tasks and $\mathcal{P}$ is the sum of all payments to primary winners. In addition, comparing line $9$ with line $17$ in algorithm listing~\ref{algo:GetWinPay} shows that the payments computed using the \emph{redundancy-reputation} factors are higher than the payments calculated by the reputation factor. Since we are concerned with minimizing the payments in general, we update the \emph{redundancy winners} in line $9$ in algorithm listing~\ref{algo:GetRedWin}).

Given the budget of the platform as in (\ref{eqn:bpv}), we can determine the remaining budget that is available--before getting a negative utility--to accomplish the tasks that have not been covered by the chosen winners \cite{b51}. Since the reputation-aware version of the proposed algorithm, \emph{RPB-RA}, results in higher payments (lines $16-18$ in algorithm listing~\ref{algo:GetWinPay}), we argue that it also motivates participants to bid for the unpopular tasks.

\begin{algorithm}
	\caption{Compute Payments for Winners}
	\label{algo:GetWinPay}
		\begin{algorithmic}[1]
			\Function{Get Winners Payments}{$V$, $B^c$, $S$, $S^{R}$, $R$, $U$}
				\For{$i \in \mathcal{I}$}
					\vspace{0.25em}
        					\State $p_i \leftarrow 0$
					\vspace{0.25em}
      				\EndFor
				\For{$i \in S \hspace{2pt} $}
					\vspace{0.25em}
        					\State $S' \leftarrow S \setminus\{i\}$, $\Theta = \Phi$
					\vspace{0.25em}
					\Repeat
						\State $q \leftarrow \argmax\limits_{\mathcal{V} \in S' \setminus \Theta} (v_\mathcal{V}^r(\Theta) - \frac{b_\mathcal{V}^c}{r_\mathcal{V}})$
						\vspace{0.25em}
						\State $p_i \leftarrow \max(p_i,\min\{v_i^r(\Theta)-(v_q^r(\Theta)-\frac{b_q^c}{r_q})\})$     				\vspace{0.0025em}
						\State $\Theta \leftarrow \Theta \cup \{q\}$
						\vspace{0.25em} 
					\Until{$\frac{b_q^c}{r_q} \geq v_q^r \bigvee \Theta = S'$}
					\vspace{0.25em}
      				\EndFor
				\vspace{0.25em}
				\For{$i \in S^{R} \hspace{2pt} $}
					\vspace{0.25em}
        					\State $S' \leftarrow S^{R} \setminus\{i\}$, $\Theta = \Phi$
					\vspace{0.25em}
					\Repeat
						\State $q \leftarrow \argmax\limits_{\mathcal{V} \in S' \setminus \Theta} (v_\mathcal{V}^{u}(\Theta) - \frac{b_\mathcal{V}^c}{u_\mathcal{V}})$
						\vspace{0.25em}
						\State $p_i \leftarrow \max(p_i,\min\{v_{i}^{u}(\Theta)-(v_q^{u}(\Theta)-\frac{b_q^c}{u_q})\})$     				\vspace{0.0025em}
						\State $\Theta \leftarrow \Theta \cup \{q\}$
						\vspace{0.25em} 
					\Until{$\frac{b_q^c}{u_q} \geq v_q^{u} \bigvee \Theta = S'$}
					\vspace{0.25em}
      				\EndFor
				\vspace{0.25em}
				\State \Return $P$
			\EndFunction
		\end{algorithmic}
\end{algorithm}

The proposed algorithm so far has not adopted the descriptive bidding of \cite{b50} and \cite{b51} at any of its stages. Basically, it selects the primary winners, then determines the redundancy winners, and then computes the payments for both. The type of bidding provided by both sets of winners is collective, and the auction management according to the aforementioned algorithm listings is reputation aware. Hence, we would compare (in terms of the attained clearance rates) the proposed procedure (up to this point) to \emph{TSCM}. However, in order to include the \emph{2SB-RA} of \cite{b50} and \cite{b51} in the comparison, one more stage should be added to the pipeline, namely, the secondary winners selection, which adopts the descriptive bidding approach.

\subsection{Secondary Winners Selection}
\label{subsec:SecWinSel} 
Unless the set of $M$ tasks have been covered by the primary and redundancy winners, the platform proceeds to the final stage of the algorithm. Using the descriptive bids proposed \textbf{\emph{in our own recent work}} \cite{b50,b51}\footnote{The code is publicly avialable through: \url{bitbucket.org/isl_aast/descriptive-bidding-ccnc-2019/src/master/}}, the platform determines another set of winners, called the secondary winners $S^{s}$, to whom the uncovered tasks are allocated. This is shown by algorithm listing~\ref{algo:SecWinSecPay} which resembles that listing provided in \cite{b51} except that the former takes $S^R$ into consideration. On the expense of the budget, the platform pays the secondary winners according to their descriptive bids in order to achieve a higher CR. This takes place after trying to cover all tasks by collective bidding through primary and redundant winners. Unlike primary and redundant winners, while assigning uncovered tasks to secondary winners, the platform ensures that a task would not be covered more than once, for the sake of better budget management.

\begin{algorithm}
	\caption{Secondary Winners Selection and Payment}
	\label{algo:SecWinSecPay}
		\begin{algorithmic}[1]
			\State $\mathcal{B} = \mathcal{V} - \mathcal{P}$
			\vspace{0.25em}
			\If{$S_{tasks} \neq T$}
				\State $S^{s} = \Phi$
				\State $T^s = \Phi$
				\For{$i \in \mathcal{I} \setminus \{S \cup S^{R}\} $} 
					\vspace{0.25em}
					\For{$t \in T_i$}
						\If{$t \in S_{tasks}$}
							\State $T_i = T_i \setminus t$
						\ElsIf{$t \in T \setminus S_{tasks}$}
							\State $P^{s} \leftarrow P^{s} \cup \{i\}$
						\EndIf
					\EndFor	
				\EndFor
			
				\vspace{0.25em}
				\State $h \leftarrow \argmax\limits_{i \in \mathcal{I}^{s}} (v_i^r(S^{s})-\frac{b_i(S^{s})}{r_i})$
				\vspace{0.25em}
				\While {$\frac{b_h}{r_h} + (r_h \times \mathcal{B}) \geq 0\hspace{2pt}\bigwedge \hspace{2pt} S^{s} \neq P^{s}\hspace{2pt} \bigwedge \hspace{2pt} S_{tasks}\neq T$}
					\vspace{0.25em}
					\vspace{0.25em}
					\State $S_{tasks} \leftarrow S_{tasks} \cup T^s_h(S^s)$
					\vspace{0.25em}
					\State $S^{s} \leftarrow S^{s} \cup \{h\}$
					\vspace{0.25em}
				 	\For{$i \in \mathcal{I}^{s} \setminus S^{s}$} 
						\vspace{0.25em}
						\For{$t \in T_i$} 
						\vspace{0.25em}
							\If{$t \in T^s_h$}
							\State $T_i = T_i \setminus t$
							\EndIf
						\EndFor
					\EndFor
					\vspace{0.25em}
					\State $\mathcal{B} = (\mathcal{B} \times r_h)-\frac{b_h}{r_h}$
					\State $h \leftarrow \argmax\limits_{i \in \mathcal{I}^{s} \setminus S^{s}} (v_i^r(S^{s})-\frac{b_i(S^{s})}{r_i})$
				\EndWhile
			\EndIf
			\State $Outlier\_Detection(S,S^{R},S^{s})$
			\For{ $s \in \{S \cup S^{R} \cup S^{s}\}$} 
				\State update $r_s$
			\EndFor
			\State \Return $S^{s}$
		\end{algorithmic}
\end{algorithm}

\section{Results and Discussion}
\label{sec:results}
In this section, \emph{design choices were made such that the values of the parameters are either identical or close to their values in other research in the literature} \cite{b43,b17,b51}, to facilitate the comparison. All simulations were done using Matlab\textsuperscript{\textregistered} 2015, on a PC with Intel Core-i7 2GHz processor and 4GB of RAM.

The simulation is done in an area of ($1000$ $m$ $\times$ $1000$ $m$) in which participants and tasks are uniformly distributed. Each participant is surrounded by an area of interest of $30m$ radius as depicted in Fig.~\ref{fig:ParticipantTask}. The value of each task and the participants' collective bids vary uniformly in [1,5] and [1,10] respectively. Similarly, the per-task bids vary uniformly in the range [$v_j - \alpha$, $v_j + \alpha$], and we set $\alpha =2$ in our simulations. The participants' reputations are varied uniformly from $0.6$ to $0.9$. We also mapped the redundancy factor to the range [0.5, 1] in order to be close to the range of the reputation to have nearly equal influence. For evaluating the effectiveness of the proposed algorithm, the \emph{RPB-RA} algorithm, we compare its performance to two algorithms from the literature, namely, \emph{Two-stage bidding (2SB)} \cite{b50,b51} and \emph{TSCM} \cite{b17} as representatives of reputation-aware techniques, \textbf{\emph{all of which are online techniques}}, i.e., they require an established connectivity between the platform and the participants. The \emph{2SB} algorithm consists of only 3 stages: 1) Primary Winners Selection, 2) Primary Winners Payment Determination, 3) Secondary Winners Selection and Payment. The redundancy stage, which is the \emph{second} stage in the \emph{RPB-RA}, is not included in it. As will be discussed below, comparing \emph{2SB} to \emph{RPB-RA} shows the impact of involving the redundancy as a factor in managing auctions. Three aspects are considered (allowed to vary) in our simulations which are: the number of auctions, the number of tasks, and the number of participants. Table~\ref{tab2} summarizes the simulated scenarios and their corresponding parameter values.

\begin{table*}[t]
\begin{center}
\caption{A summary of the different simulated scenarios and their corresponding parameter values.}
\setlength{\abovecaptionskip}{5pt plus 0pt minus 0pt}
\setlength{\belowcaptionskip}{-10pt plus 0pt minus 0pt}
\includegraphics[width=6in]{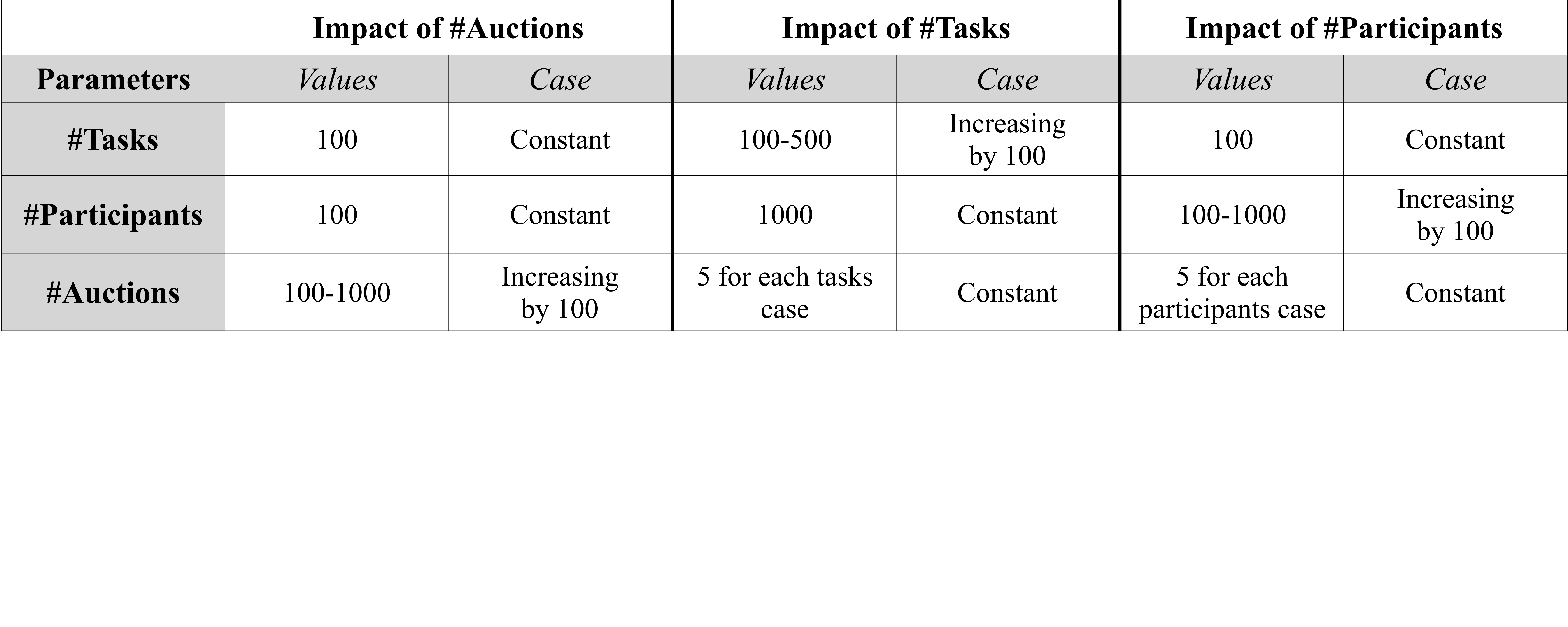}
\label{tab2}
\end{center}
\end{table*}

\subsection{The impact of varying the number of auctions on the CR}
\begin{figure}[t]
\begin{center}
\setlength{\abovecaptionskip}{5pt plus 0pt minus 0pt}
\setlength{\belowcaptionskip}{-10pt plus 0pt minus 0pt}
\includegraphics[width=3.4in]{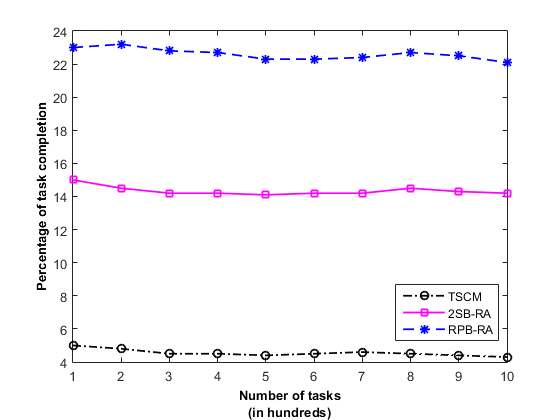}
\caption{The impact of varying the number of auctions on the performance of different reputation-aware incentive mechanisms.}
\label{fig:first}
\end{center}
\end{figure}

First, we investigate the effect of varying the number of held auctions on the performance of the aforementioned algorithms. Fig.~\ref{fig:first} shows that our algorithm results in a significant increase in clearance rate, that is close to five times that of the \emph{TSCM} and two times that of \emph{2SB}. The average percentage of tasks completion is nearly constant, regardless the number of held auctions. 

\subsection{The impact of varying the number of tasks on the CR}
\begin{figure}[t]
\begin{center}
\setlength{\abovecaptionskip}{5pt plus 0pt minus 0pt}
\setlength{\belowcaptionskip}{-10pt plus 0pt minus 0pt}
\includegraphics[width=3.4in]{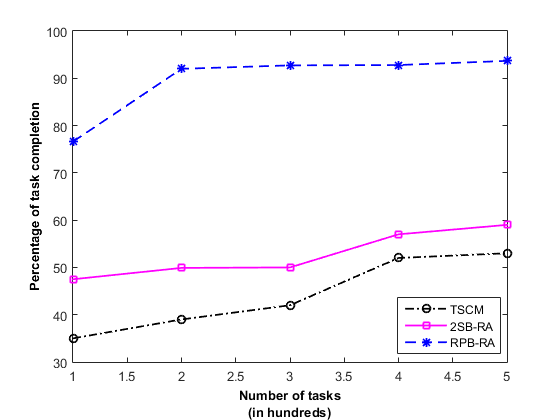}
\caption{The impact of varying the number of tasks on the performance of different reputation aware incentive mechanisms}
\label{fig:third}
\end{center}
\end{figure}

In Fig.~\ref{fig:third}, we compare the performance of the three algorithms, with regards to the achieved CR, under different number of tasks. In addition, we assume that there are 100 participants in the area of interest who are ready to share their sensing data. From Fig. ~\ref{fig:third}, we can see that the CR always increases with the increasing number of tasks, because more tasks are added in the area of interest--of the participants--so the winners perform more tasks and the CR increases accordingly. It is clear that the \emph{RPB-RA} outperforms the \emph{TSCM} and \emph{2SB} across a wide range of task number. The CR slightly exceeds \emph{90\%} in case the number of tasks slightly exceeds $200$ tasks. 

We justify the significant increase in CR by highlighting the fact that other techniques aim at maximizing the user and the platform utility using only one stage of bidding (collective bidding). However, towards the goal of optimizing the utilization of the budget and better satisfy service demanders, our algorithm realizes three winning stages throughout the pipeline, with corresponding three types of winners. It is important to mention that \emph{our algorithm does not increase the budget of the platform, but it uses it more efficiently and economically}. Instead of leaving the unpopular tasks uncovered till the stage of selecting secondary winners as in \emph{2SB}, it covers these tasks first. Hence, it pays for the redundancy winners using the collective bids which is cheaper than the descriptive bids as discussed in \cite{b50,b51}.

\subsection{The impact of varying the number of participants on the CR}
\begin{figure}[t]
	\begin{center}
		\setlength{\abovecaptionskip}{5pt plus 0pt minus 0pt}
		\setlength{\belowcaptionskip}{-10pt plus 0pt minus 0pt}
		\includegraphics[width=3.4in]{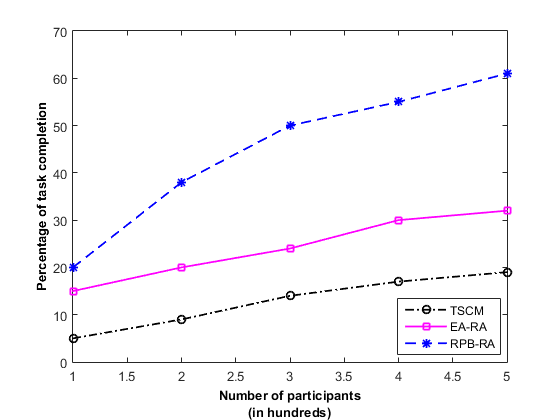}
		\caption{The impact of varying the number of participants on the performance of RPB-RA.}
		\label{fig:second}
	\end{center}
\end{figure}

As shown in Fig.~\ref{fig:second}, when the number of participants increases, more candidates compete to be chosen by the platform. Hence, the probability of finding a better choice of winners, among the newly added participants, increases. Thus, the CR increases. Our proposed method attains consistently higher CR, though, compared to the other techniques. This increase is approximately four times the CR of \emph{TSCM} and almost linear in the range of $100-500$ participants.

\section{Conclusion}
\label{sec:conclusion}
This research is concerned with enhancing the quality of task allocation in participatory auction-based MCS. We proposed a new bidding procedure (\emph{RPB-RA}) that maximizes the CR by assigning higher priority to the unpopular sensing tasks in a campaign, i.e., the tasks that are expected to be completed by a fewer number of bidders. Our own previous work addressed the maximization of CR using descriptive bidding. However, to the best of our knowledge, this research is the first to address the maximization of CR through penalizing redundant task assignment. The free parameters of the proposed algorithm were identified and we simulated varying scenarios (varying number of auctions, tasks, and participants) in order to evaluate that framework. We showed that \emph{RPB-RA} outperforms the state-of-the-art approaches, with regards to the CR. For future work, we will consider other factors that may affect the selection of participants in multi-task MCS environments. New optimization methods and theoretical models for the platform utility will also be studied.

\end{document}